\documentclass{article}

\usepackage{arxiv}

\usepackage[utf8]{inputenc} 
\usepackage[T1]{fontenc}    
\usepackage{hyperref}       
\usepackage{url}            
\usepackage{booktabs}       
\usepackage{amsfonts}       
\usepackage{nicefrac}       
\usepackage{microtype}      
\usepackage{graphicx}
\usepackage{natbib}
\usepackage{doi}

\title{Identification of crowds using mobile crowd detection (MCS) and visualization with the DBSCAN algorithm for a Smart Campus environment
}

\author{ \href{https://orcid.org/0009-0006-1317-5362}{\includegraphics[scale=0.06]{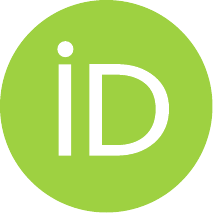}\hspace{1mm}Luis ~Chirinos-Apaza}\\
	Faculty of Statistic and Computer Science\\
	National University of Altiplano\\
	Avenida floral 1153, Puno, Peru \\
	\texttt{lchirinosa@epg.unap.edu.pe} \\
}

\renewcommand{\headeright}{}
\renewcommand{\undertitle}{}
\renewcommand{\shorttitle}{}

\hypersetup{
pdftitle={Identification of crowds using mobile crowd detection (MCS) and visualization with the DBSCAN algorithm for a Smart Campus environment},
pdfsubject={cs.DC},
pdfauthor={Luis ~Chirinos-Apaza},
pdfkeywords={Crowd detection,
Mobile Crowd Sensing (MCS),
DBSCAN,
University campus,
Security,
Public space management,
Simulation,
Mobile application},
}

\begin{document}
\maketitle

\begin{abstract}
 Multidisciplinary research, in conjunction with artificial intelligence (AI), the Internet of Things (IoT), Blockchain and Big Data analysis, has lowered barriers and made companies more productive, in other words, the joint work of these areas has promoted digital transformation in all areas, for example Artificial intelligence (AI) has made it possible to automate processes, and the Internet of Things (IoT) has connected devices and physical objects, enabling real-time data collection and analysis. Blockchain has provided a secure and transparent way to transact and store data. Big Data analysis has allowed companies to obtain valuable insights from large amounts of data. As these technologies continue to evolve, we can expect to see even more innovations and benefits in the future. This paper explores the feasibility of using Mobile Crowd Sensing (MCS) and visualization algorithms to detect crowding on a university campus. A survey was conducted to evaluate the university community's perception of a mobile application that provides information about crowds, and a detection scenario was simulated using randomly generated data and the DBSCAN algorithm for visualization. Preliminary results suggest that the system is viable and could be a useful tool for the prevention of accidents due to crowding and for the management of public spaces. The limitations of the study are discussed and future lines of research are proposed, such as crowd prediction, data privacy, and visualization optimization.

\end{abstract}

\keywords{Crowd detection \and 
Mobile Crowd Sensing (MCS) \and 
DBSCAN \and 
University campus \and 
Security \and 
Public space management \and 
Simulation \and 
Mobile application}

\section{Introduction}
Multidisciplinary research in conjunction with artificial intelligence, with IoT, Blockchain and Big Data Analyst has reduced barriers and made businesses more productive, that is, the joint work of these areas has promoted digital transformation \cite{makkar_handbook_2024} wherever we go, but it's not just about business, digital transformation has turned traditional cities into smart cities, improving the lives of citizens. which is the main objective of a smart city, another combination of areas such as machine learning, data mining and video processing equally contribute to the transformation from a normal city to a smart city \cite{el-alfy_intelligent_2023}, It is as if there were multiple trajectories to travel and encounter various obstacles that deserve to be resolved while traveling these paths. The final objective of this research is identical: to contribute to the development of the city and, therefore, of society.

The prediction models as made by \cite{wang_real-time_2023} allow the detection of crowds of people on city streets, offering valuable information for urban management and public safety.

The incorporation of these technologies (computer vision, machine learning) in an intelligent urban environment has shown positive results, although it requires a significant economic investment in the necessary equipment and its maintenance, as well as in specialized technical personnel for its operation. Cameras, sensors, and Wi-Fi networks have contributed significantly to the security and management of spaces or crowds, but they will not be the only leading technologies in this, and we will develop that in this research.

Despite significant progress in the development of smart cities in the world's major urban centers, Peruvian cities have not received the same level of attention. In Peru, progress in this area has been limited, despite the evident need for greater participation in the implementation of smart solutions to address pressing challenges, such as accident prevention and organized crime.

The motivation for researching alternative and lower-cost solutions for the preventive detection of crowds of people anywhere on the university campus arose from the belief that a smart campus acts as a minimalist version of a city. intelligent \cite{polin_making_2023}, \cite{negreiros_smart_2020}. Therefore, investigating first in a small space, such as a university campus, will allow us to obtain valuable information for later application in larger environments, such as the cities of Peru. Therefore, this preliminary research addresses its viability to later be applied to the city and be a high-impact investigation.

Given the challenges we have identified, we do not intend to replace existing solutions from technologies such as computer vision or pattern recognition with cameras and sensors. Instead, we propose a meaningful complement to these technologies. Furthermore, for cities that do not have a crowd detection system implemented due to difficulties in investment, maintenance, permits, or operating licenses, we suggest as a precedent a form of crowd detection based on Mobile Crowd Sensing (MCS) and algorithms. visualization such as DBSCAN or even heat map visualization. We consider these methods appropriate for identifying and visualizing where excessive crowds are generating.

In this research, we aim to explore the feasibility of leveraging MCS and visualization algorithms to assist citizens and authorities in the prevention and management of crowds. We believe these approaches could be valuable in notifying citizens about potential crowding hazards and helping authorities make better decisions about how to prevent and manage public spaces in the face of potential uncontrolled crowd growth and in order to carry out this research, we will need to collect data on the location and movement of people. One way to do this is to develop a mobile application that allows users to report their location and movement. Another option is to generate random data that represents the behavior of people in a public space and once we have the data, we will use MCS and visualization algorithms to develop a system that can detect and track crowds of people. This system could then be used to notify citizens about potential crowding hazards and to help authorities make better decisions about how to prevent and manage public spaces in the face of potential uncontrolled crowd growth.

To validate the foundations of our research, we carried out a survey among students, teachers, and administrative staff to find out their perception of a mobile application proposal. This application would provide information about dangerous crowds, with incentives to increase its usefulness and allow managers to visualize information about crowds and streamline their decision-making process.

We have quickly reviewed the literature on how MCS, data visualization, data privacy, and prediction are used on the global stage, including Peru, although with few publications in the country.

\section{Related work}
Our priority search engine is Scopus, one of the largest repositories of scientific research articles worldwide. The following searches gave us a wide discovery of new concepts such as MCS. Our query searches are:
\begin{itemize}
    \item TITLE-ABS-KEY("crowd management" AND "smart cities") AND ( LIMIT-TO ( SUBJAREA,"COMP" ) OR LIMIT-TO ( SUBJAREA,"ENGI" ))
    \item TITLE-ABS-KEY ( "peru" AND ( "smart city" OR "smart campus" ) AND "machine learning" )
    \item TITLE-ABS-KEY ( ( "information" OR "data" OR "datasets" ) AND ( "federated learning" OR "edge computing" OR "IoT" OR "Internet of Things" ) AND ( "machine learning" OR "deep learning" ) AND ( "decentralization" OR "ICT" OR "security" OR "smart city environment" OR "smart campus" ) ) AND PUBYEAR > 2020 ) 
    \item TITLE-ABS-KEY(("BIG DATA") AND ("SMART CITY" OR "SMART CAMPUS")) AND ( LIMIT-TO ( PUBYEAR,2024) OR LIMIT-TO ( PUBYEAR,2023) OR LIMIT-TO ( PUBYEAR,2022)
\end{itemize}

Studies like that of \cite{bamaqa_simcd_2022} indicate that intelligent crowd management solutions can prevent crowding disasters through efficient crowd learning models; these solutions involve monitoring crowds and modeling their dynamics. Crowd datasets can be real or synthetic. Real data are expensive and difficult to acquire, so simulation tools are used to generate synthetic datasets. The author presents the process of generating synthetic crowd datasets for crowd anomaly detection and prediction, the data sets include two types of crowd anomalies: SIMulated Crowd Data (SIMCD)-Single Anomaly and SIMCD-Multiple Anomalies for crowd prediction and anomaly detection models.

The research of \cite{zhang_forecasting_2022} proposes a model to predict city-wide crowds at fine spatiotemporal scale; it integrates a Gated Recurrent Unit (GRU), a convolutional neural network, and k-nearest neighbors, achieving a better accuracy and lower training time cost than existing models.

The application created by the authors \cite{dong_oncampus_2016}, they suggest three services: commerce, forum and educational information that will not only help the university campus, but will facilitate the adaptation and start of the construction of a Smart Campus environment. This application is called OnCampus, proving useful to the university community and satisfying the well-being of the population. 

An investigation based on decision making such as that of \cite{doorley_revurb_2019} analyzes dense public spaces that host social activities and that are part of a successful urban district proposes to examine the relationship between physical characteristics and functions in said urban environments, for this purpose they use a Non-Homogeneous Poisson process generating a positive correlation between stores, entertainment , parking and bus stops. 

CrowdTelescope is a prediction framework developed by \cite{zhang_crowdtelescope_2023} which performs crowd flow prediction using neural networks and spatiotemporal graphs, efficiently collects a dataset of WiFi connections on a university campus. 

A Crowdsensing system implemented, in a mobile application, turns mobile phones into monitors for the detection of weather data and air pollution, using CNN, LSTM, improving portability and accuracy \cite{liu_third-eye_2018}.

A recent investigation by \cite{madyatmadja_classifying_2023} handles the term Crowdsourcing, which means outsourcing collaboration services to distributed groups. Taking into account the LAKSA application, it classifies citizen complaints using data mining and studies in K-Nearest Neighborgs, Random Forest, Super Vector Machine and AdaBoost for the prediction evaluation. 

A Research that left us interested is \cite{aboualola_edge_2023}, where they mention that smart devices can improve security in smart cities; technologies such as edge sensing, IoT, big data and AI can be used through smart devices to create emergency-aware systems. It also conducts a review of the literature of the disaster and emergency management, highlighting the role of edge technologies and the importance of social media and AI.

The study carried out by \cite{pacheco_smart_2024} analyzes the viability of an evaluation model for smart cities in Lima, Peru. 80 municipal officials were surveyed. The results revealed gaps in information availability, energy conservation and educational level. Transformation processes promote development, quality of life and citizen equality. The study contributes to a better understanding to build more sustainable cities.

One of the investigations that encouraged us to start with the smart campus concept instead of smart city was \cite{fortes_campus_2019} where he explains how Internet of Things (IoT) seeks to connect objects to the network to efficiently monitor and manage the environment. IoT applications have been limited by computational capacity and communications efficiency, but new technologies make it possible to overcome these problems. The University of Malaga is committed to researching and innovating the Smart-Campus concept, transforming university campuses into "small" smart cities. This article presents the commitment of the University of Malaga to the development of its Smart-Campus in the fields of its infrastructure, management, research support and learning activities.

A classic investigation of \cite{lau_extracting_2017} gives us confidence in the direction of our study, smartphones with sensors allow us to take advantage of various information thanks to the crowd detection mobile application. Traditional approaches have drawbacks such as high battery consumption, to mitigate this, they proposed an undersampling point of interest (POI) extraction framework based on stay point detection (VSPD) and sensor fusion-based environment classification (SFEC). The Density Based Spatial Clustering Algorithm with Noise (DBSCAN) produces the most accurate result. The SFEC model classifies the indoor or outdoor environment of the grouped POI, real-world data demonstrate the effectiveness of the low sampling rate model on high-noise spatial-temporal data.

CrowdSenSim is a simulator, created by \cite{fiandrino_crowdsensim_2017}, for mobile crowdsensing designed for urban environments and smart city services. It evaluates the performance of MCS systems in different sensing paradigms (participatory and opportunistic), is applicable to smart street lighting scenarios and shows a visualization using Google Heatmap Tool.

According to \cite{ray_survey_2023}, smart devices, such as smartphones, can sense and communicate with their environments. This capability has been used to create ubiquitous-computing applications. Mobile crowd sensing is a new paradigm that uses sensors on smart devices to collect data from the environment and transmit them to the cloud for analysis. Crowdsourcing is a model for solving complex, distributed problems in nature using a crowd of indeterminate size. Smart devices with the ability to sense the environment and utilize the wisdom of the crowd can be used for various benefits of society, for a better standard of living.

Authors like \cite{capponi_survey_2019} highlight that MCS has attracted attention in recent years and has become an attractive paradigm for urban sensing. MCS systems are based on the contribution of mobile devices from a large number of participants, or a crowd. Despite growing interest in the research community, MCS solutions need deeper investigation and categorization of many aspects. They present a study on existing works in the domain and propose a detailed taxonomy to shed light on the current panorama and classify applications, methodologies, and architectures.

The author of \cite{ferreira_predicting_2023} proposes an approach to predict the concentration and movements of people in a smart city. Taking advantage of mobile phone data, the author develops a predictive model that shows the ability to predict areas of high concentration of people. The approach respects user privacy and was evaluated using real data from Lisbon; the results demonstrate the precision and effectiveness of the approach.

The article published by the authors of \cite{pereira_analysis_2022} analyzes two machine-learning algorithms, DBSCAN and Local Outlier Factor (LOF), in detecting outliers in a continuous framework for point-of-interest (PoI) detection. With two datasets, LOF exhibits the best performance, making it most suitable for outlier detection in near-real-time PoI detection.


A framework, developed by \cite{yu_crowdkit_2024}, details how Mobile Crowdsensing (MCS) allows participants to use mobile devices to share information related to a common interest. Typical applications include environmental monitoring, intelligent transportation systems, and public safety. Existing frameworks for MCS applications are specific and lack extensibility, which makes innovation difficult. CrowdKit is a generic, developer-oriented programming framework that abstracts common data models and MCS application functions, which makes them reusable. It follows the principles of modular design, visual development, and automatic code generation to provide extensibility and reduce the difficulty and time cost for developers. Its algorithm modules can accommodate various, advanced MCS algorithms. CrowdKit is simple, general, extensible, and highly efficient.

\section{Methodology}
In the present study, we will adopt a mixed research approach combining quantitative and qualitative methods. On the one hand, we will carry out surveys on a representative sample of the student population of the National University of the Altiplano Puno. On the other hand, we will carry out a review of the academic literature on Mobile Crowd Sensing (MCS) that has been used for the detection of crowds in urban environments. Visualization algorithms, such as DBSCAN, which provide a graphical representation of the data, will also be analyzed and collected.

In addition, we will simulate a crowd detection scenario based on randomly generated datasets in a specific geographic area, such as the university campus. We will visualize and classify the results of this simulation to identify potential crowding problems. Finally, the surveys carried out will allow us to evaluate the feasibility and interest of a sector of the university community in the development of solutions to address the challenges associated with the agglomeration of people.

\subsection{Survey development}
We carried out a survey aimed at a group of students and staff who attend the university campus, which consisted of 10 questions that was created by the Google Forms platform shown in Table \ref{tab:asks}.

\begin{table}[ht]
	\caption{List of questions asked for the survey}
	\centering
	\begin{tabular}{p{0.5cm} p{0.9\linewidth}}
		\toprule
		\cmidrule(r){1-2}
		 N° & Questions  \\
		\midrule
		1 & You attend the university in person because you play the role of:  \\
		2 & How often do you get around the city on foot?   \\
		3 & Do you think that being in the middle of a growing and uncontrolled crowd could be a problem? \\
            4 & Have you experienced crowds in the city or on campus? \\
            5 &  Would you use an app to avoid dangerous crowds? \\
            6 & What additional functions would you add to a crowd information app? \\
            7 & Are you worried that a mobile app is collecting your location data? \\
            8 &  Would you share your anonymous location data to improve campus or city security and management? \\
            9 & Would integrating incentives into a mobile app improve its level of adoption and usage? \\
            10 & What type of incentive would you like the mobile application to have for you to use it? \\
		\bottomrule
	\end{tabular}
	\label{tab:asks}
\end{table}

\subsection{Collection of data}
As part of the research plan, the development of a mobile application similar to the proposals made by the authors is proposed. \cite{dong_oncampus_2016}, \cite{wirz_coenosense_2013}, \cite{bujari_mobile_2020}. 

The purpose of this application is to collect relevant data from users which facilitates the MCS approach. This approach allows for similar effectiveness to the implementation of Internet of Things (IoT) cameras and sensors, but at a lower cost; this is because users become our sensors to detect crowds.

However, for this mobile application and its use to fulfill its objective, it is essential that the user is well informed about the capabilities of MCS and has completely acceptable concerns about their data; for that reason, in the survey, we included a question to find out in which situations users would give their consent to share their positioning data, in order to encourage the user to use the mobile application and participate in the research.

According to various investigations of the state of the art, such as \cite{lendak_mobile_2016}, the future challenges of this line of research are promising; another is \cite{fiandrino_crowdsensim_2017}, where many more sensors of mobile devices are considered, such as microphones, cameras, accelerometers, barometers or gyroscopes. The characteristics of the data collected from our users are shown in Table \ref{tab:table}:

\begin{table}
	\caption{Features of data collected from users}
	\centering
	\begin{tabular}{lll}
		\toprule
		\cmidrule(r){1-3}
		Name     & Description     & Data Type \\
		\midrule
		ID &  Integer & Unique number that identifies each data point      \\
		  Longitude   & Float   & Geographic longitude in decimal degrees      \\
		Latitude     & Float & Geographic latitude in decimal degrees     \\
            Timestamp  &  Datetime & Timestamp at the time of activating data sending  \\
		\bottomrule
	\end{tabular}
	\label{tab:table}
\end{table}

\subsection{Processing}
It is essential to ensure data consistency and the conformity of each feature to the type of data it will represent for the dataset. It would be highly advantageous to achieve a mobile application implementation for our data collection, even for new business ideas, such as question number 6 of the survey carried out. However, due to limited time for this preliminary investigation, we will program a python script to randomly generate geographic points within the university campus area to continue this study.

\subsection{Clustering}
We consider the DBSCAN algorithm an ideal tool for this type of point clustering due to its ability to group densely populated points based on their parameters. Specifically, the $epsilon$ parameter defines the distance that one point must be from another to be considered within the group, and the $min\_points$ parameter establishes the number of points that can surround a point. This feature makes it a superior option in this situation, because the distribution of points, or people's locations, on the plane is completely irregular. Unlike the k-means clustering algorithm, DBSCAN does not require points to follow ordered patterns; rather, they can be freely positioned and moved, making it difficult to apply the k-means algorithm in this context.

In this study, we mainly evaluate and contrast the DBSCAN clustering algorithm, using the available datasets; if we find inconsistencies in these algorithms, we will propose a new proposal to improve the precision and efficiency of pattern identification in dense spaces.

\subsection{Implementation}
\begin{figure}
	\centering
	\fbox{\includegraphics[width=0.8\linewidth]{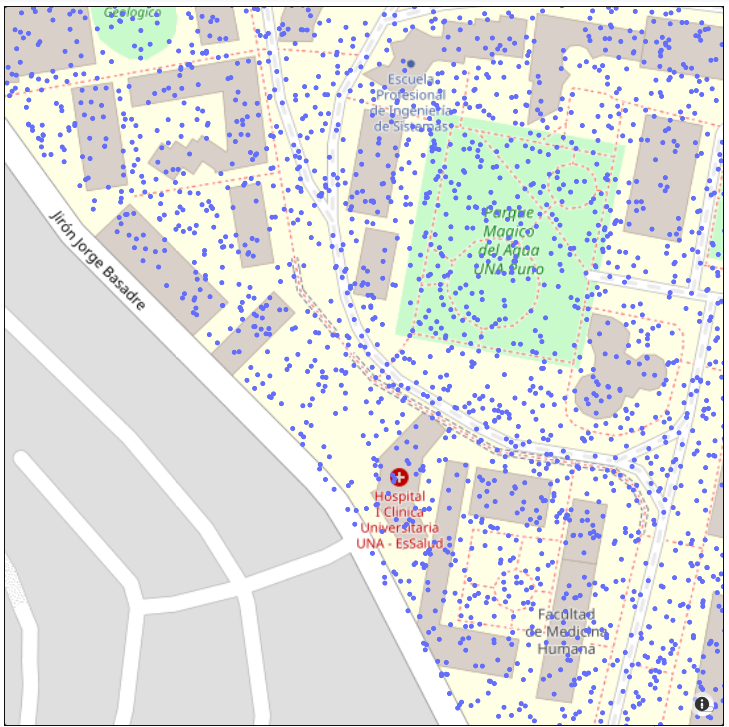}}
	\caption{Observation of the randomly generated locations until the development of the application for better data samples, the points blues representing people within the university campus captured at a specific hour and minute.}
	\label{fig:map1}
\end{figure}

\begin{figure}
	\centering
	\fbox{\includegraphics[width=0.8\linewidth]{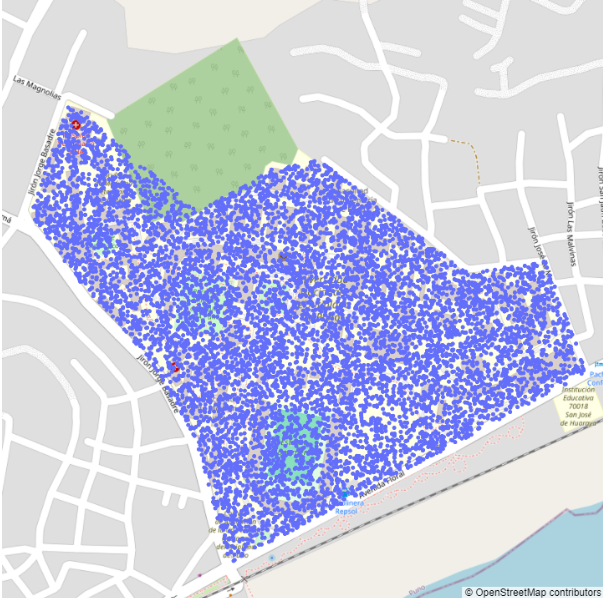}}
	\caption{View of the university campus under study with blue dots that represent the number of people who entered the campus at a specific time.}
	\label{fig:plot1}
\end{figure}

Based on data generated from the university's field locations and its typical capacity to accommodate students, we estimate that about 50\% of the student population could attend campus; this is equivalent to approximately 9,000 people within the university campus, according to public information from the government transparency portal. Please note that this estimated figure may be subject to change in the future to reflect a more realistic environment; However, this represents a useful initial simulation for our current purposes, shown in figures \ref{fig:map1} and \ref{fig:plot1}.

\subsection{Display of agglomerations}
When visualizing crowds of people, we turn to visualization libraries, such as Seaborn and Matplotlib and Pandas for reading the datasets; these tools allow us to generate heat maps and other essential graphics to visualize areas with a higher density of people; this information is vital for making quick and effective decisions in urban environments, such as cities or university campuses.

\begin{figure}
	\centering
	\fbox{\includegraphics[width=0.8\linewidth]{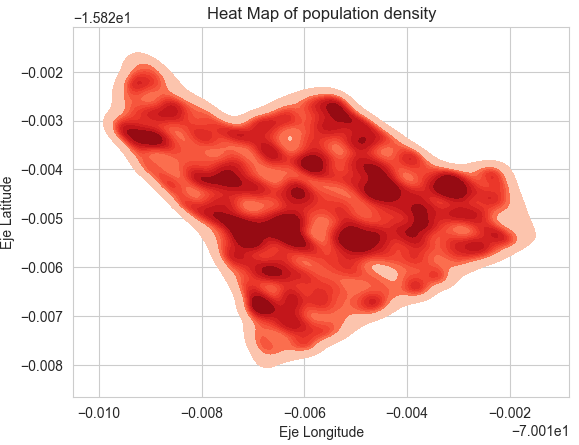}}
	\caption{Simple visualization implemented with the Seaborn library on the datasets and we can visualize the dark areas that indicate greater agglomeration of points in space.}
	\label{fig:plot2}
\end{figure}

In the figure, we see an example of visualization using a heat map of the data density. In this type of visualization, shown in Figure \ref{fig:plot2}, darker colors indicate more crowding of data, while lighter colors indicate less crowding. Heat maps are a useful tool to communicate clear information about the current situation and facilitate decision-making.

Heatmaps are an effective tool for identifying patterns and trends in data; they can help identify areas of high and low density, as well as areas of change or activity; they can also be used to compare different datasets and to identify correlations between variables. Heatmaps are a valuable tool for data visualization; they can help communicate clear information about the current situation and facilitate decision-making.

\begin{figure}
	\centering
	\fbox{\includegraphics[width=0.8\linewidth]{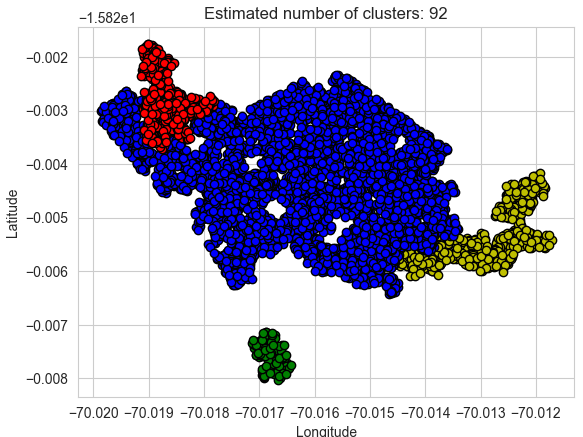}}
	\caption{Visualization of the points using the \texttt{sklearn.cluster DBSCAN} library where the parameters were: $eps=0.0007$ and $min\_samples=4$}
	\label{fig:plot3}
\end{figure}

Due to the nature of our generated dataset, we experienced difficulties in visualizing the number of point groups when using the DBSCAN algorithm; however, as seen in Figure \ref{fig:plot3}, we consider that the selection of the $eps$ and $min\_samples$ parameters was the most optimal, and helped us to clearly differentiate several well-defined and specific agglomerations in the plane.

\subsection{Decision making and preventive actions}
Decision-making is crucial in the university environment, since it empowers us to adopt preventive measures against potential risk situations that may occur on campus. Among the multiple circumstances that may arise, the high concentration of people affects the well-being of passersby, as evidenced in the survey in Table \ref{tab:asks}, where the majority of respondents claim to have experienced worrying situations when finding themselves in the middle of a crowd. Scenarios, such as long waiting lines to access the university dining room or arriving at a study room and finding it completely occupied, mean delays and loss of time in most cases, forcing students to look for other study rooms without knowing whether they will be empty or full. This situation highlights the relevance of our research and its practical application in our university environment, and we presume that this is also the case in other, similar environments. Crowd visualization is very useful for space management or security personnel, because it helps them take preventive measures.

\section{Survey results}
\begin{figure}
	\centering
	\fbox{\includegraphics[width=0.4\linewidth]{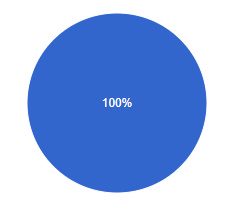}}
	\caption{Unanimous response given to the question of whether they ever felt in a fairly crowded situation.}
	\label{fig:s1}
\end{figure}

\begin{figure}
	\centering
	\fbox{\includegraphics[width=0.4\linewidth]{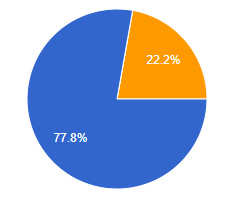}}
	\caption{Response where 77.8\% responded that they would use the application if it did not have security problems and was full of incentives and 22.2\% responded that they might use the app.}
	\label{fig:s2}
\end{figure}

The results of the survey allow us to confirm that our research is consistent with our reality and is aimed at contributing to future solutions in this regard, as shown in Figures \ref{fig:s1} and \ref{fig:s2}.

The survey has provided us with valuable information that will help us better understand the problem, and it is still active so far, with 90 participants, and we are hoping the number of participants continue to increase for a better sample. At the moment, these are some of the key findings from the survey include:

\begin{itemize}
    \item The majority of respondents are concerned about the problem of crowding.
    \item Most respondents are willing to use the app if the data protection guarantee disappears
    \item The majority of respondents are willing to use the application if it brings incentives
    \item Respondents prefer to have first-hand information about what alternative routes they can choose to avoid congested areas.
\end{itemize}

Other findings found in the survey suggest that incentives to use the application are:

\begin{itemize}
    \item Economic incentives
    \item The exact time that the next-closest public transport bus will arrive to your location
\end{itemize}

We are grateful to everyone who participated in the survey; your answers will help us make a difference.

\section{Discussions}
With the results obtained, the DBSCAN algorithm has been optimal to help us visualize clusters of dense points on the map; however, a comparison with other visualization algorithms is necessary, as well as an evaluation of the performance of the algorithms. Despite the preliminary nature of this research, it is the line of research that we are interested in carrying out, given that we consider that there are compelling reasons for carrying it out. We trust that the results obtained will be well-received, due to their intention to be a study not frequently carried out in our local society; even compared to the pace of other cities, we consider it relevant and a great precedent for a possible transition from a traditional city to a smart city. We do not intend to replace other, widely studied methodologies, such as computer vision or pattern recognition and the use of cameras and sensors; on the contrary, this line of research in MCS will very well complement the concerns we have regarding the high agglomeration of people and the decision-making of an authority or manager.

The survey provides us with evidence of this, and indicates that our proposal is viable. The objective is to contribute to society and reduce problems, such as those related to accident prevention, improvements in safety, better management of public space, and effective communication.

An aspect that has not been mentioned at the beginning, regarding the types of data that will serve as the basis for the dataset, are the other variables, such as weather and holidays, which expands our panorama of the amount of data that we must process. 

Future lines of research that we open with this research will be discussed in future works.

\section{Conclusions}

We can conclude that the research on Mobile Crowd Sensing (MCS) and clustering models for a Smart Campus environment, despite using a simulated dataset, provides us with a realistic and credible result on how we can realize of the formation of dangerous crowds and make preventive decisions accordingly.

It is a methodology that complements the detection of crowds of people through cameras and sensors; this technology promises to be very useful for both the university community and administrators when making decisions related to accident prevention, safety, and community well-being; furthermore, this research provides us with a solid precedent to extend our work and apply it to the city as a whole, which will be our long-term goal. Likewise, we are opening the way for the implementation of these technologies in a practical solution, such as a mobile application with innovative features, as demonstrated by the results of the survey carried out.

\section{Future works}
In the context of our focus on crowd detection, we have identified a number of promising research avenues that will allow us to expand our understanding and improve our detection techniques.

First of all, we must improve the dataset by implementing a mobile application that will be essential to receive realistic data; this mobile application is mainly designed to have, as its main functionality, information about where the busiest places of people are located, in real time; the design of the application must be as visual and user-friendly as possible, and we do not rule out the inclusion of more functionalities, such as those most voted for in the survey.

One of these ways is crowd prediction, based on our detection findings, we seek to develop predictive models that allow us to anticipate the movement of the detected crowd, these models could be based on machine learning techniques or physical models of crowd behavior, by predicting crowd movement, we can improve the effectiveness of crowd management measures and reduce the risk of incidents.

An additional concern is the privacy of user data when they use the mobile device to transmit their locations, according to the user survey, they are generally very or moderately concerned about privacy, although in the end, they do not care, therefore, research in federated learning becomes fundamental and mandatory to obtain better results and greater user confidence.

Another path to explore is visualization optimization; currently, we use clustering algorithms, such as DBSCAN, to visualize the detected agglomerations; however, there are other clustering algorithms that could be helpful in expanding our knowledge; additionally, we could explore the use of alternative visualization techniques, such as heat maps or network graphs, to better communicate information about agglomerations, without leaving aside other technologies for integration, such as Blockchain, IoT sensors, social networks, meteorological data; the opportunities multiply as we continue to investigate more.

We consider it essential to investigate the scalability and performance of networks; as the number of people using our systems increases, it is essential to ensure our network infrastructure can handle the load; otherwise, we could experience a network infrastructure collapse, poor performance, or even a simulated DDoS attack; to evaluate this situation, we will perform performance tests and analyze the behavior of the network under different conditions.

These are just some of the promising research avenues we are exploring in the context of our focus on crowd detection, we believe this research will allow us to improve our understanding of crowding and develop better tools to detect and manage it.

\bibliographystyle{unsrtnat}
\bibliography{references}  






\end{document}